\documentclass[aps,pre,showpacs,superscriptaddress,twocolumn]{revtex4}
\usepackage{graphicx,color,epsfig}
\usepackage{amssymb,amsmath}

\begin{document}
\title{Against mass media trends: minority growth in cultural globalization}
\author{M. G. Cosenza}
\affiliation{School of Physical Sciences \& Nanotechnology, Universidad Yachay Tech, 100119 Urcuqu\'i, Ecuador}
\author{M. E. Gavidia}  
\affiliation{Luxembourg Centre for Systems Biomedicine, University of Luxembourg, Belvaux, Luxembourg}
\author{J. C. Gonz\'alez-Avella}
\affiliation{APSL, 07121 Parc Bit, Palma de Mallorca, Spain}

\begin{abstract}
We investigate the collective behavior of a globalized society under the influence 
of endogenous mass media trends. The mass media trend is a global field corresponding 
to the statistical mode of the states of the agents in the system. 
The interaction dynamics is based on Axelrod's rules for the dissemination 
of culture. We find situations where the largest minority group, possessing a cultural state
different from that of the predominant trend transmitted by the mass media, 
can grow to almost half of the size of the population.
We show that this phenomenon occurs when
a critical number of long-range connections are present in the underlying network of interactions.
We have numerically characterized four phases on the space of parameters of the system: 
an ordered phase;  a semi-ordered phase where almost half of the population consists of the largest 
minority in a state different from that of the mass media;  a disordered phase; 
and a chimera-like phase where one large domain coexists with many very small domains.
\end{abstract}


\maketitle

\section{Introduction}
In a social context, globalization  implies increasing interconnectivity
among people: any agent may interact with any other in a social system.
Internet, social networks, mobile phones, and other technologies
have contributed to the realization of fully connected societies.
In recent years, the globalization of culture has been an issue of much 
interest in Social Sciences
\cite{Jensen}.
It generally refers to the
exchange of cultural symbols among people around the world that leads to
the formation of shared norms and behaviors with which people associate 
their individual identities and collective culture \cite{Hopper,James4,Albrow}. 
Although there is no consensus on the consequences of globalization on 
local cultures,
many scholars sustain that the
exposure of minority groups to a global culture can undermine their own cultural 
identity \cite{Castells,Ritzer,Croisy,Kaul}.
One of the most controversial debates on the matter of cultural globalization and 
the ``minority identity crisis'' involves the role of mass media as a facilitating tool 
for its expansion or limitation \cite{Curran,Sparks,Rantanen,Gentz}.

With the advent of globalization, 
broadcasting, telecommunication, and internet-based companies have also become global; 
mass media messages can practically reach 
all individuals and groups in the society.
The interaction with mass media 
have experienced important changes:   
in on one hand people get informed by the media, while on the other hand
the information is influenced by the evolution of people's preferences \cite{James1,Crane,Mccombs}. 
To some extent, main stream media adapt their contents to
reflect the predominant tendencies, fashionable behavior, and cultural trends
in the globalized society \cite{Knut}. 
Mass media trends represent a plurality
information feedback in social systems with endogenous
cultural influences \cite{Shibanai,JASSS,Plos}. 
Global information feedback also occurs, for example, in 
virtual platforms, such as Amazon, where users have access to 
the average valuation of a commercial 
product resulting from their own opinions. 

Models of mass media acting on social systems 
have been shown to produce counterintuitive effects: cultural diversity 
increases when the intensity of the mass media message is strong enough, 
whereas weaker messages impose their state leading to global cultural homogeneity 
\cite{GCT,Global,Candia,NJP,Peres}.

In this article we investigate the collective behavior of a globalized society under the influence 
of endogenous mass media trends. 
As a model for this system, we consider a fully connected network 
of social agents subject to a global field representing the mass media trend. 
In particular, we study the evolution of minority groups 
in presence of the predominant cultural trends transmitted by the media.
We employ Axelrod's rules for the dissemination 
of culture as interaction dynamics  \cite{Axelrod}, a non-equilibrium model of wide interest 
for physicists \cite{Castellano,Vilone,Klemm,Maxi,GCT,Global,Candia,NJP,Peres,Kuper,Galla,Floria}.
We find situations where the predominant tendencies of mass media are not followed by the entire population. 
We uncover a nontrivial minority growth phenomenon
induced by the endogenous mass media for some values of parameters:
a minority group 
possessing a cultural 
state alternative to that of the mass media trend can increase its size up to
almost half of the size of the population. 
Additionally, for some conditions the system segregates
into two distinguishable subsets: agents in
one subset 
share the state of the mass media trend,
while agents in the other subset exhibit a disordered or incoherent state. 
This configuration is analogous to a chimera state arising in networks of
globally coupled oscillators.

In Sec.~II we present the model for a globally connected network of social agents
subject to mass media trends. 
In Sec.~III we
describe the growth of a minority 
and characterize the collective states
arising on the space of parameters of the system. 
The dynamics of the minority growth phenomenon is  investigated in Sec.~IV. 
Section~V contains our conclusions.

%\section{Materials and methods}
\section{Model for mass media trends in a globally connected network}
We consider a population of $N$ social agents consisting of a fully connected network,
where every agent can interact with any other in the system. We assume that interactions take place
according to the dynamics of the model of cultural dissemination of Axelrod \cite{Axelrod}. 
The state variable of agent $i$ $(i=1,2,\ldots,N)$ is given by the $F$-component vector $x_i=(x_i^1, \ldots,x_i^f, \ldots, x_i^F)$,
where each component $x_i^f$ represents a cultural feature that can take any of $q$ different traits or options in the set $\{0, 1, . . . , q-1\}$. 
In this paper we employ the normalized parameter $Q=1-(1-1/q)^F$ to represent the decreasing number of options per feature, such that $Q=0$ for $q \to \infty$ (large number of options) and $Q=1$ for $q=1$ (one option).
We denote by $M=(\mu^1,\ldots,\mu^f,\ldots,\mu^F)$ the global field defined as 
the statistical mode of the states in the system at a given time. 
Then, the component $\mu^f$ 
corresponds to the most abundant value shown by the component $x_i^f$ of all the state vectors 
$x_i$ in the population. If the most abundant value is not unique, one of the possibilities
is chosen at random with equal probability. 
We assume that each agent is subject to the influence of the global field $M$.
In the context of cultural models, this global field represents an endogeneous mass media influence acting on
all the agents in the system and containing the most predominant trait in each cultural feature
present in the population; i. e., a global cultural trend or fashionable behavior.

In this paper we fix the parameter value $F=10$. 
Initially,
the states $x_i$ are assigned at random with a uniform distribution. 
At any given time, a randomly selected agent 
can interact either with any other agent in the system or with the global field $M$,
according to the rules of Axelrod's cultural model.
We define the dynamics of the system by the following iterative algorithm:

\begin{enumerate}
  \item Select at random an agent $i$, called \textit{active agent}.
  \item Select the \textit{source of interaction}: with probability $B$ active agent $i$ interacts 
  with the field $M$, while with probability $(1-B)$ it interacts with another agent $j$ chosen at random.
  \item Calculate the overlap between the active agent and the source of interaction, 
  defined as the number of shared components between their respective vector states, 
  $l(i,y) = \sum_{f=1}^{F} \delta_{x_i^f,y^f}$,
  where $y^f=\mu^f$ if the source of interaction is the field $M$, or $y^f=x_j^f$ if $i$ interacts with $j$.
  We use the delta Kronecker function: $\delta_{x,y} = 1$, if $x=y$; $\delta_{x,y} = 0$, if $x\neq y$.
 \item  If $0<l(i,y)<F$, with probability $l(i,y)/F$,
 choose $h$ such that $x_i^h \neq y^h$ and set $x_i^h = y^h$;
if $l(i,y) = 0$ or $l(i,y) = F$ the state of agent $i$ does not change.
  \item If the  source of interaction is $M$,  update the field $M$.
\end{enumerate}

The parameter $B \in [0,1]$ in step (2) describes
the probability of interaction of the agents with the mass media message and represents the
intensity of the field $M$. Steps (3) and (4) describe the interaction rules from Axelrod's model.
Step (5) characterizes the time scale for the updating of the global field $M$. 
It expresses the assumption that agents do not have instantaneous information about the 
the mass media trend, but only when they effectively interact
with the media.
As the intensity $B$ of the global field increases, 
the updating rate of the state of the global field also increases.

In the absence of mass media influence ($B=0$), a system subject to Axelrod's dynamics
reaches an asymptotic state in any finite network where domains of different sizes are formed.
A domain is a set of connected
agents that share the same state. A homogeneous collective state or ordered phase in the system is characterized by having overlap $l(i,j) = F$, $\forall i,j$.
This state possesses $q^F$ equivalent realizations. In an inhomogeneous collective state or disordered phase, several domains
coexist. It is known that, on several networks, the system reaches
an ordered phase for values $q < q_c$ , and a disordered phase for $q > q_c$,
where $q_c$ is a critical point \cite{Castellano,Vilone,Klemm,Fede}. 
In terms of the normalized parameter $Q$, the disordered phase occurs for $Q < Q_c = 1-(1-1 /q_c)^F$ and the ordered phase arises for $Q > Q_c$.

%\section{Results}
\section{Emergence of minorities and chimera states}

To characterize the collective behavior of the system under the influence of global mass media trends, 
we employ two statistical quantities as order parameters:
(i) the average normalized size (divided by $N$) of the largest domain, denoted by $S_1$; and (ii) the average normalized size 
of the second largest domain, assigned as $S_2$. 
% It has been shown that, without mass media influence,
% the critical value $q_c$ for the order-disorder transition on 
% a fully connected network  
% depends on the system size as $q_c\sim N$ \cite{Fede}.
Thus, in the absence of mass media influence ($B=0$), 
an ordered phase for values $Q > Q_c$ is characterized by $S_1 \to 1$, while
a disordered phase for $Q < Q_c$ has $S_1 \to 0$.
% 
% the system reaches an ordered phase characterized by $S_1 \to 1$ for values $Q > Q_c$, 
% and a disordered phase with $S_1 \to 0$, for $Q < Q_c$. 

\begin{figure}[h]
\includegraphics[scale=.3]{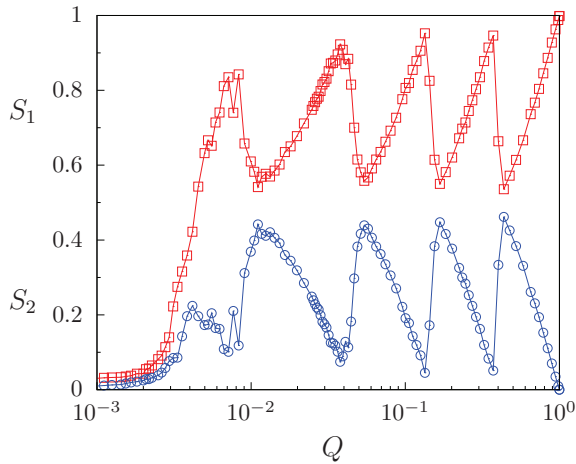}
\caption{Order parameters $S_1$ (squares) and $S_2$ (circles) as functions of $Q$ (log scale)
for $B = 0.8$.
Each data point is the result of averaging over $100$ realizations of initial conditions.
Fixed parameters are $F=10$, $N=800$.}
\label{f1}
\end{figure}

As the probability $B$ increases from zero, the behavior of the quantity $S_1$
notably changes.
In Fig.~\ref{f1} we show both order parameters $S_1$ and $S_2$ as functions
of $Q$ for a fixed value of $B>0$.
The quantity $S_1$ exhibits several local minima 
with $S_1 <1$ at some values $Q >Q_c$; we have found that
both these values and the number of minima depend on $B$.
For values $Q<Q_c$, the system settles into a disordered state 
where $S_1 \to 0$.

The state of the largest domain always corresponds to the state of the field $M$ that represents
the predominant cultural trend. 
The presence of 
local minima in $S_1$ indicates that, for some values of parameters, 
the cultural trend is not adopted by all 
the agents in the population. 
Figure~\ref{f1} shows 
that the local minima of $S_1$ are 
associated to local maximum values of $S_2$, 
such that $S_1+S_2 \approx 1$ for $Q >Q_c$. Therefore, two large domains that occupy almost
all the system can arise 
for $Q >Q_c$.
The state of the largest domain is equal to that of the field $M$, but 
the second largest domain reaches a state different from $M$.
Thus, the values of $Q$
for which $S_1$ displays local minima are related to the emergence of the largest minority group 
ordered in a state alternative (non-interacting) to that being imposed by the mass media trend. 

\begin{figure}[h]
\includegraphics[scale=.3]{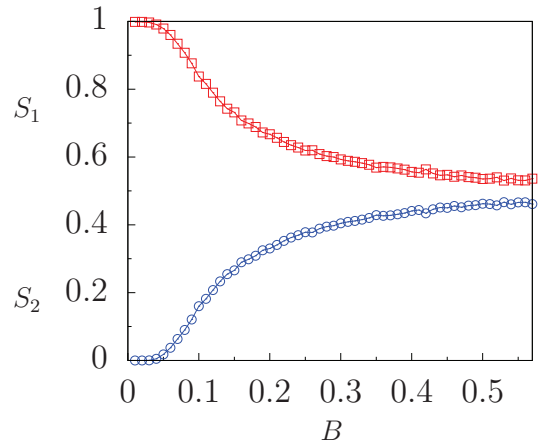}
\caption{Quantities $S_1$ (squares) and $S_2$ (circles) as functions of $B$, with fixed $Q=0.40$ ($q=20$). 
Each data point is the result of averaging over $100$ realizations of initial conditions.
Fixed parameters: $F=10$, $N=800$.}
\label{f2}
\end{figure}

To investigate the influence of the mass media cultural trend on the behavior of the largest 
minority in the population, 
in Fig.~\ref{f2} we show both  quantities $S_1$ and $S_2$ as functions of $B$ for a fixed value
of parameter $Q >Q_c$ where
a local minimum of $S_1$ appears.
For small values of $B$, the largest domain makes up all the system, and thus $S_1=1$, $S_2=0$.
As $B$ increases, a competition takes place between the spontaneous order emerging in the system
due to the agent-agent interactions
and the order being imposed by the endogenous global field. As a consequence, a second largest 
domain emerges and grows its size 
while the largest domain decreases its size, such that both constitute almost all the population,
$S_1+S_2 \simeq 1$. Thus, we have the counter-intuitive result that, for some values of $Q$ 
that represent the number of options per cultural feature,
mass media transmitting the
predominant cultural trend can actually promote the growth of the largest minority group up to almost
half of the population size in a state different to that of the media. We denote this situation
as a minority-growth state.

For greater values of $B$, the interaction of the mass media field with the agents dominates, and $S_1$
may increase. 
The case $B=1$ describes a situation where the agents solely interact with the global field $M$, which becomes 
initially established and then remains fixed. This scenario is indistinguishable from that of a system subject to a fixed external 
global field \cite{NJP}.
Then, the fraction of agents whose states converge to the state of the field is the fraction of agents that 
initially share at least one component with the field and it is given by \cite{NJP}
\begin{equation}
S_1(B=1)=1-(1-1/q)^F=Q.
\label{B1}
\end{equation}

\begin{figure}[h]
\includegraphics[scale=.35]{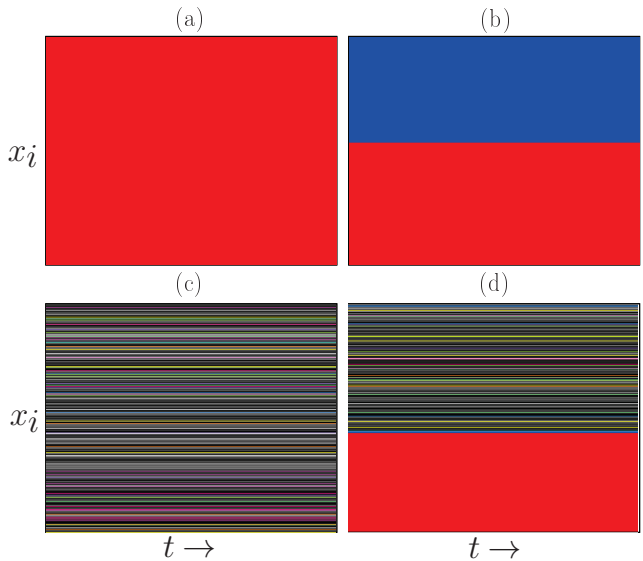}
\caption{Asymptotic states $x_i$ ($i=1,\ldots,N$) of the agents (vertical axis) as a function of time (horizontal axis) over $500$ time steps,
for different parameter values $B$ and $q$. Each value of the state variable of an
agent is represented by a different color; if $x_i=x_j$, then $i$ and $j$ share the same color. 
Fixed parameters:  $F=10$, $N=800$.
(a) $B = 0.1$, $Q=0.79$ (homogeneous state, phase I); (b) $B = 0.9$, $Q=0.01$ (minority-growth state, 
phase II); 
(c) $B =0.03$, $Q=0.0019$ (disordered state, phase III); (d) $B = 0.48$, $Q=0.0037$ (chimera state, phase IV).}
\label{f3}
%$q = 7$ $q = 1400$ ($q = 5300$) ($q = 2700$)
\end{figure}

In addition to the above described collective states, for some parameter values $B$ and $Q >Q_c$, the system 
can reach a partially coherent configuration where one part of the population forms a domain in a homogeneous state equal to 
$M$, while the remaining part is in a disordered state. This state is characterized by $S_1 >0$ and $S_2 \to 0$; the size of $S_1$ depends 
on initial conditions. This situation is similar to a chimera state occurring in dynamical systems
subject to global interactions, where the symmetry of the system is spontaneously broken into
two coexisting coherent (or synchronized) and incoherent (or desynchronized) 
subsets \cite{Kaneko1,Sen,Pik,Schmidt,Mis}. Chimera states where initially identified
in systems of nonlocally coupled oscillators \cite{Kuramoto,Abrams}. 
Figure~\ref{f3} displays the asymptotic spatiotemporal patterns corresponding to 
the collective behaviors observed in the system 
subject to the influence of the global mass media trend, for different values of parameters $B$ and $Q$. 

\begin{figure}[h]
\includegraphics[scale=.65]{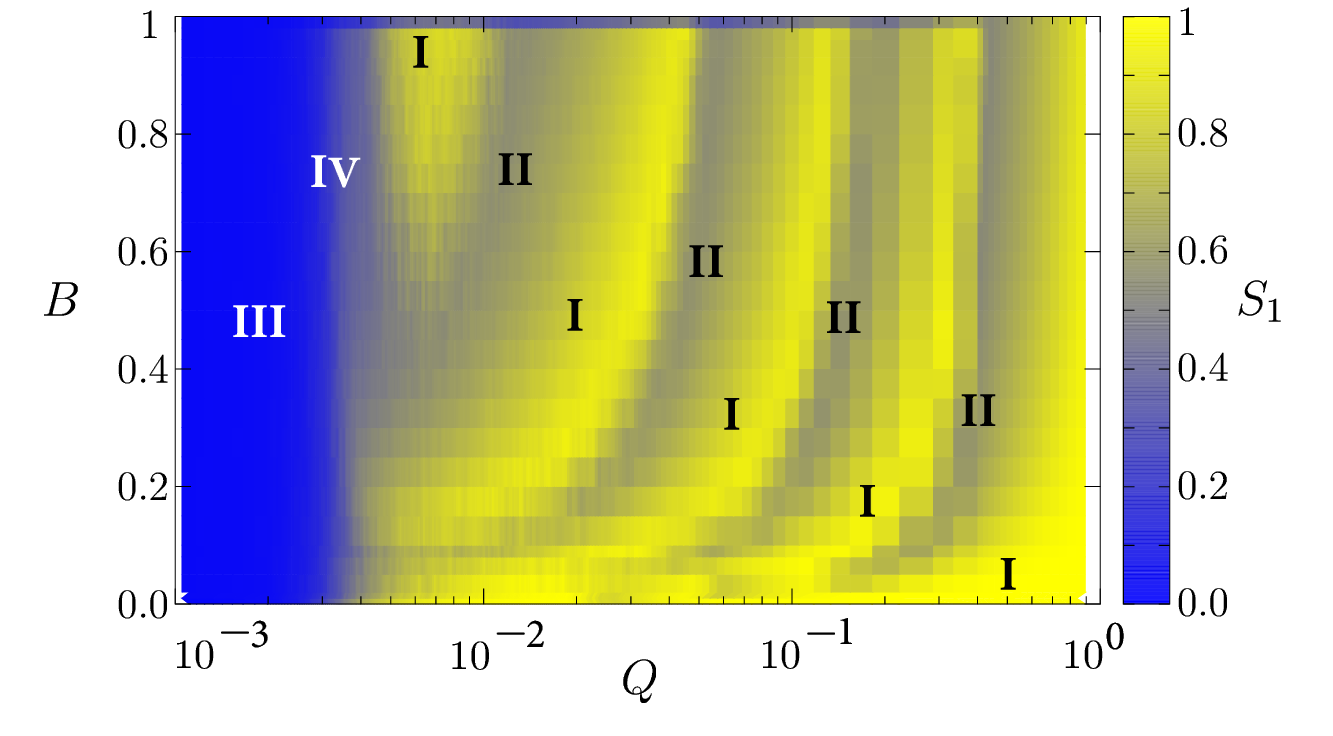}
\caption{Phase diagram for the system on the space of parameters $(B,Q)$.
The color code represents the value of the order parameter $S_1$. The regions
where the different phases occur are labeled: phase I,  homogeneous state;
phase II, minority-growth state; phase III, disordered state; phase IV, chimera-like states.
Fixed parameters are $F=10$, $N=800$. 
Each data point is averaged over $100$ realizations of initial conditions.}
\label{f4} 
\end{figure}

The collective behavior of the system can be characterized on the space of parameters
$(B,Q)$, as shown in Fig.~\ref{f4}. Four phases can be found: (I) a homogeneous, ordered phase for which $S_1 \to 1$;
(II) a semi-ordered phase, where $S_2 >0$  and $S_1+S_2 \simeq 1$, characterized by the growth of a second largest domain
ordered in a state different from that of the mass media; (III) a disordered phase for $Q < Q_c$, for which $S_1 \to 0$;
and (IV) chimera states, characterized by $S_1 >0$ and $S_2 \to 0$,
where one large domain coexists with 
many domains of negligible sizes. 
The chimera states depend on initial conditions, and they 
can emerge in regions of parameters that lie between phase I and phase III states. In fact, chimera states share
features of both phase I and phase III; they can be considered as transition configurations
between these two phases.

\section{Dynamics of minority growth}
\begin{figure}[h]
\includegraphics[scale=.3]{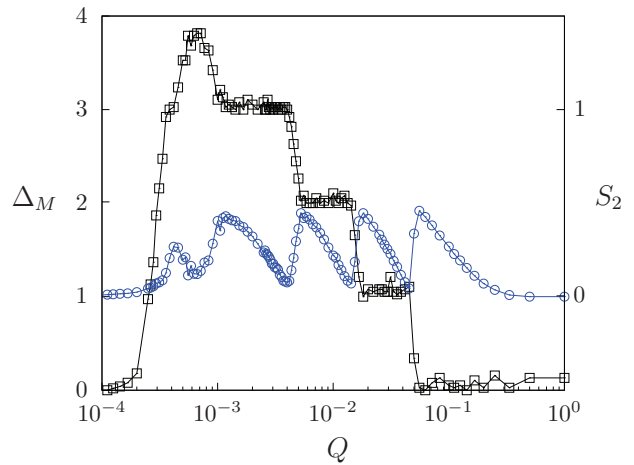}
\caption{Average number of updates $\Delta M$ of the mass media vector (squares) as a function of $Q$ 
(log scale), with fixed $B=0.8$.
Each value $\Delta M$ is averaged over $50$ realizations of initial conditions.
For reference, the graph $S_2$ versus $Q$ (circles) with  $B=0.8$ is included without vertical scale.
Fixed parameters $F=10$, $N=800$.}
\label{f5}
\end{figure}

As the system evolves from its initial conditions for given parameter values, 
the mass media vector $M$ changes
until it reaches a stationary state.
In Fig.~\ref{f5} we calculate the average number of updates or
changes $\Delta M$ occurred to vector $M$ as a function of $Q$, for a fixed value of the intensity $B=0.8$. 
For comparison, the curve $S_2$ (right vertical scale) versus $Q$ 
for this value of $B$ from Fig.~\ref{f1} 
is also included.

For values $Q < Q_c$, there is a large number of domains and vector $M$ adopts the state of any
of them with equal probability since these states are equally abundant. Then, 
$M$ does not change, and $\Delta M=0$. At the critical point $Q=Q_c$ there occur many fluctuations
in the number of domains and their sizes. As a consequence, vector $M$ experiences several changes
and $\Delta M$ suddenly increases to a value of almost $4$. 
Since there is a
probability $1-B$ that
 all agents interact among themselves to form domains,
the order emerging from the agent-agent
interactions competes with the order being imposed by $M$ for $Q >Q_c$.
A  change in $M$ represents a reset
of initial conditions, and thus is equivalent to a change in the number of agents that can interact
with the updated vector $M$, a number reflected in the size $S_1$. 
The competition between the  agent-field  and agent-agent interactions
takes place over 
a range of values of $Q$ where the number of changes $\Delta M$ remains constant,
allowing for the increase of $S_1$ and a decrease of $S_2$.    
Increasing $Q$ implies less available states and a reduction in the number of changes $\Delta M$.
A drop in $\Delta M$ 
represents a decrease in $S_1$ and gives chance to
the rise of the minority size $S_2$.
These processes are repeated  on a sequence of intervals of  $Q$, until this parameter
 approaches the value $Q=1$  where there are few states available and
 the global field does not change, $\Delta M=0$; 
 the agents mostly interact with a fixed global field $M$ that imposes its state on them. 
Thus, the interplay between the 
global coupling among the agents and the adaptive dynamics of the autonomous global field $M$
is responsible for the repeated occurrence of 
minority growth in a state alternative to that of the field as $Q$ is varied.

\begin{figure}[h]
\includegraphics[scale=.3]{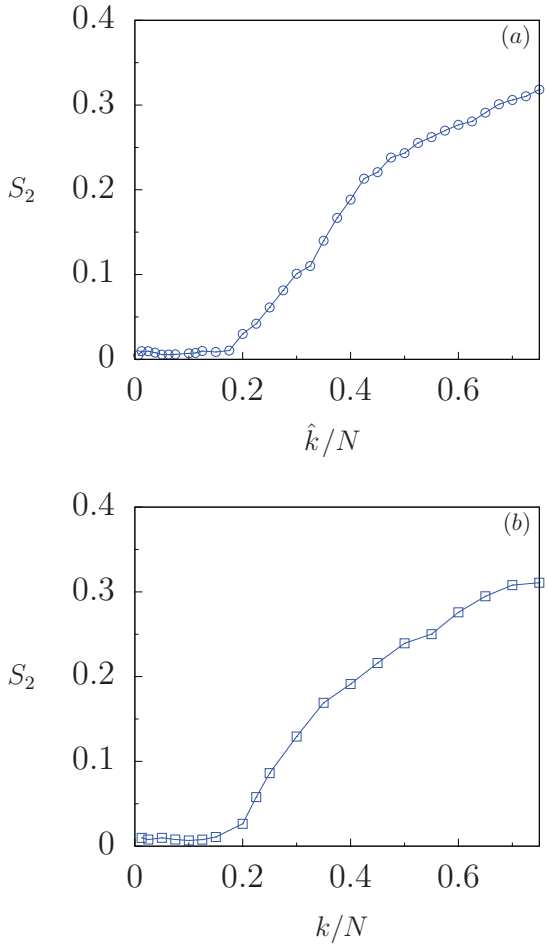}
\caption{(a) $S_2$ as a function of the average number of neighbors $\bar k$ for the system defined on a random network.
(b) $S_2$ as a function of the number of neighbors $k$ for the system defined on a ring with long range interactions.
Fixed parameters in both cases are $Q=0.247$, $B = 0.01$, $N=800$.}
%$q = 400$
\label{f6}
\end{figure}

To elucidate the role of the
connectivity of the network on the minority growth phenomenon, 
we consider the dynamics of the system defined on two different topologies: 
(i) an Erd\"os-Renyi random network of $N$ nodes having average degree $\bar k$; 
and (ii) a ring of $N$ nodes where each node is connected to $k$ neighbors, $k/2$ on each side.
The fully connected network studied above corresponds to the values $\bar k = N-1$ and $k=N-1$, respectively.
Figure~\ref{f6}(a) shows the size of the second largest domain $S_2$ as a function of $\bar k$ 
for random networks, with
fixed values of $B$ and $Q>Q_c$. When $\bar k$  is small, the system reaches a homogeneous 
state with $S_2=0$ and $S_1=1$. 
However, $S_2$ grows as the average number 
of neighbors increases above a threshold value $\bar k \approx 0.23N$. Similarly, Fig.~\ref{f6}(b) shows $S_2$ 
as a function of $k$ for a ring-type network. The size $S_2$ increases from zero above some
threshold value of the number of neighbors $k\approx 0.23N$. 
Thus, all-to-all interactions are not essential for the growth of a minority group in a state different
from that of the global mass media
trend on a network; rather this effect appears when a critical range of interaction is reached,
which is a small 
fraction of the size of the system.

\begin{figure}[h]
\includegraphics[scale=.29]{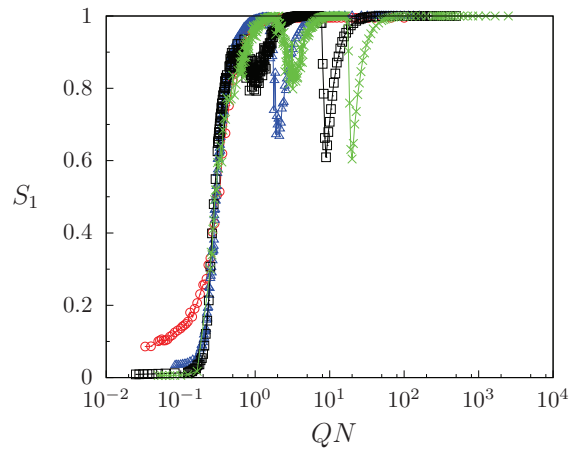}
\caption{$S_1$ as a function of $QN$ (log scale) with fixed $B=0.01$ for different sizes of the system: $N=200$
(circles), $N=800$ (triangles), $N=2500$ (squares), and 
$N=5000$ (crosses). Each data point is averaged over $100$ realizations of initial conditions.
Fixed parameter $F=10$.}
\label{f7}
\end{figure}

Finally, we have explored the behavior of the system for different population sizes $N$. Figure~\ref{f7}
shows the quantity $S_1$ as a
function of $QN$ with a fixed mass media intensity $B$, for different values of $N$.
The critical point for the transition to phase III scales as $Q_c \sim N^{-1}$, as expected 
for a fully connected network \cite{Fede}.
We have verified that phases I, II, II, and IV 
continue to form as $N$ increases. 
However, the drop in size of  $S_1$, and therefore 
the rise of the largest minority group $S_2$, are both enhanced with increasing system size.

\section{Conclusion}

We have addressed the question of the competition between
the collective self-organization of a globalized society and the influence 
of endogenous mass media trends, as well as the role of the network
connectivity in this competition.  The mass media is a global field corresponding 
to the statistical mode of the states of the social agents. 
As interaction dynamics, we have considered Axelrod's rules for the dissemination 
of culture, a non-equilibrium model whose dynamics allows the existence of non-interacting states.
By studying this model on a fully connected network, we have found 
a phenomenon of spontaneous growth of the largest
minority domain in state non-interacting with that of the mass media trend; this domain can
occupy  almost half of the size of the system. 
The growth of a largest minority group reflects
the tendency towards the spontaneous order related to the agent-agent interactions.
We have shown that the competition between the global coupling
among the agents and the adaptive nature of the mass media field leads to the repetition of the raise of
the largest minority for several values of parameters.

We have characterized four phases for the collective behavior of the system on its space of parameters: 
(I) an ordered phase; (II) a semi-ordered phase where almost half of the system consists of the largest 
minority in a state different from that of the mass media; (III) a disordered phase; 
and (IV) a chimera-like phase where one large domain coexists with many very small domains.

By considering a random network with varying average degree and a network with varying range of interactions,
we have shown that the occurrence of growth of the largest minority against the mass media trend  
is related to the presence of a critical number of long-range connections in the underlying network. 
Thus, all-to-all interactions are not necessary for observing this phenomenon.

The phenomenon of minority growth in the presence of an autonomous global field should also expected in
other non-equilibrium systems having non-interacting states, such as
social and biological systems whose dynamics usually possess a
bound condition or threshold for interaction. This involves several opinion models as well as 
models of motile agents
such as swarms, bird flocks, fish shoals, bacteria colonies, and non-local interactions in population dynamics.

\section*{Acknowledgments}
This work was supported  by Corporaci\'on Ecuatoriana para el Desarrollo de la Investigaci\'on 
y Academia (CEDIA) through project CEPRA-XIII-2019 ``Sistemas Complejos''.

\end{document}